\documentstyle[12pt,epsfig]{article}
\newcommand{\be}{\begin{equation}}
\newcommand{\ee}{\end{equation}}
\newcommand{\ba}{\begin{eqnarray}}
\newcommand{\ea}{\end{eqnarray}}
\newcommand{\baa}{\begin{eqnarray*}}
\newcommand{\eaa}{\end{eqnarray*}}
\newcommand{\bb}{\begin{thebibliography}}
\newcommand{\eb}{\end{thebibliography}}

\newcommand{\ci}[1]{\cite{#1}}
\newcommand{\bi}[1]{\bibitem{#1}}
\begin{document}
\begin{flushright}
CPT-2002/P.4352 \\
July 2002
\end{flushright}

\bigskip\bigskip

\begin{center}
{\Large \bf Bjorken Sum Rule at low $Q^2$}\\[1cm]
{Jacques Soffer}\\[0.3cm]
{\it Centre de Physique Th\'eorique - CNRS - Luminy,\\
Case 907 F-13288 Marseille Cedex 9 - France}
\footnote{E-mail: soffer@cpt.univ-mrs.fr}
\\[0.3cm]
and\\[0.3cm]
{Oleg V. Teryaev}\\[0.3cm]
{\it Bogoliubov Laboratory of Theoretical Physics\\
Joint Institute for Nuclear Research, Dubna\\
Head Post Office, P.O. Box 79, 101000 Moscow, Russia}
\footnote{E--mail: teryaev@thsun1.jinr.dubna.su}\\[1.2cm]
\end{center}
\begin{abstract}
A description of the generalized Gerasimov-Drell-Hearn
sum rules for proton and neutron is suggested, using
their relation to the
Bjorken sum rule. The results support an earlier conjecture, that the
structure function $g_T$ features a smooth $Q^2-$dependence,
while the structure function $g_2$ is changing rapidly,
due to the elastic contribution to the Burkhardt-Cottingham sum rule.
A possible violation of this later sum rule is briefly discussed.

\end{abstract}

\newpage

The Bjorken sum rule \cite{Bj} is known to be one of the most fundamental
constraints for our understanding of the nucleon spin structure. Its appearance
in QCD relies on the Operator Product Expansion, describing the large
$Q^2$ region, while at moderate $Q^2$ the perturbative \cite{Pert}
and power \cite{Power} corrections are of major importance.
Moreover, the transition to the entirely non-perturbative $Q^2$ region is
rather cumbersome.

Nevertheless, the limit for $Q^2=0$ is provided by taking
the difference for proton and neutron of
the Gerasimov-Drell-Hearn (GDH) sum rules \ci{Ger,DH}.
Contrary to the proton case, for this difference $p-n$, the asymptotic and
real photon values are of the same sign which provides the opportunity
for checking a possible smooth approximation \ci{book}.
In this paper we study this extrapolation in some detail.

The generalized ($Q^2$-dependent) GDH sum rules are
just being tested experimentally with a high accuracy for both proton and
neutron\cite{E143,HER,ZEM}.
The striking feature of the proton data is the low ($\sim 200-250 MeV^2$)
``crossover'' point,  which is in complete agreement
with our prediction \cite{ST93,ST95}, published almost 10 years ago.
Our approach is
making use of the relation to the Burkhardt-Cottingham
sum rule for structure function $g_2$, whose elastic contribution
is the main source of a strong $Q^2-$dependence, while
the contribution of the other structure function, $g_T=g_1+g_2$ is smooth.

However, the preliminary neutron data \cite{ZEM} are going well above
the prediction, made in the similar manner \cite{ST97}.
We address this problem and show, that
the reason for the discrepancy is the model for the neutron
structure function $g_T^n$.
In fact, the general arguments of \cite{ST95} are not
applicable for the neutron, and the model used in Ref.\cite{ST97}
appears to be more or less {\it ad hoc}.

To avoid this problem, we consider here the $Q^2$-dependence
of the non-singlet combination  $g_1^p-g_1^n$,
whose asymptotic behaviour is described by the Bjorken sum rule.
We apply the same method and describe its behaviour in the low $Q^2$
region, together with its proton and neutron components, being
in a reasonable agreement with recent experimental data.

To recall our approach let us first note, that the presence of
$g_2$ in the description of the longitudinal polarization is by
no means surprising, as soon as one uses the language of invariant,
rather than helicity amplitudes \cite{ST95}.

To define the spin-dependent structure functions one should express
the antisymmetric part of the hadronic tensor $W^{\mu \nu}$ as a
linear combination of all possible Lorentz-covariant tensors. These
tensors should be orthogonal to the virtual photon momentum $q$, as
required by gauge invariance, and they are linear in the nucleon
covariant polarization $s$, from a general property
of the density matrix. If the nucleon has momentum $p$, we have as
usual,
$s\cdot p=0$ and $s^2=-1$. There are only two such tensors: the first one
arises already in the Born diagram
\be T_1^{\mu
\nu}= \epsilon^{\mu \nu \alpha \beta}
s_\alpha q_\beta
\ee
and the second
tensor is just
\be
T_2^{\mu \nu}=(s\cdot q) \epsilon^{\mu \nu \alpha\beta}p_{\alpha}q_{\beta}.
\ee
The scalar coefficients of these tensors are specified in a
well-known way, since we have
\ba
W^{\mu \nu}_A={-i\epsilon^
{\mu \nu  \alpha\beta}\over {p\cdot q}}q_{\beta}
(g_1 (x, Q^2)s_{\alpha}+g_2 (x,Q^2)(s_{\alpha}-p_{\alpha}{s\cdot q \over
{p\cdot q}}))=\nonumber \\
{-i\epsilon^ {\mu \nu  \alpha\beta}\over {p\cdot q}}q_{\beta}((g_1 (x,
Q^2)+g_2 (x,Q^2))s_{\alpha}-g_2
(x,Q^2)p_{\alpha}{s\cdot q \over {p\cdot q}})~. 
\ea
This tells us that $g_2$, due to the factor $(s\cdot q)$, is making the difference
between longitudinal and transverse polarizations, while
$g_T=g_1+g_2$ contributes equally in both cases.
More exactly, $g_2$ provides this difference because
its contribution is
non-zero in the case of longitudinal polarization and is zero in the case
of transverse polarization. It is just in this sense that we are speaking
about the $g_2$ contribution to longitudinal polarization.

Let us consider the $Q^2$-dependent integral
\be
I_1(Q^2)={2 M^2\over {Q^2}} \int^1_0 g_1(x,Q^2) dx~.
\label{I1}
\ee
It is defined for {\it all} $Q^2$, and $g_1(x,Q^2)$ is the obvious
generalization for all $Q^2$ of the standard scale-invariant $g_1(x)$.
Note that the elastic contribution at $x=1$ is not
included in the above sum rule. Then one recovers at $Q^2=0$
the GDH sum rule
\be
I_1(0)=-{\mu_A^2 \over 4}
\ee
where $\mu_A$ is the nucleon anomalous magnetic moment in nuclear magnetons.
While $I_1(0)$ is always negative, its value at large $Q^2$ is determined
by the $Q^2$ independent integral $\int^1_0 g_1(x) dx$, which is
positive for the proton and negative for the neutron.

The separation of the contributions of $g_T$ and $g_2$
leads to the decomposition of  $I_1(Q^2)$ as the
difference between $I_{T}(Q^2)$ and $I_2(Q^2)$
\be
I_1(Q^2)=I_{T}(Q^2)-I_2(Q^2),
\ee
where
\be
I_{T}(Q^2)={2 M^2\over {Q^2}} \int^1_0 g_{T}(x,Q^2) dx,
\;\;\;\;I_2(Q^2)={2 M^2\over {Q^2}} \int^1_0 g_2(x,Q^2) dx~.
\ee

There are solid theoretical arguments to expect a strong $Q^2$-dependence
of $I_2(Q^2)$. It is the well-known Burkhardt-Cottingham sum rule \ci{BC},
derived independently by Schwinger \ci{Sch},
using a rather different method. It states that
\be
\label{el}
I_2(Q^2)={1\over 4}\mu G_M (Q^2)
\frac{\mu G_M (Q^2) - G_E (Q^2)}{1+\frac{Q^2}{4M^2}},
\ee
where $\mu$ is the nucleon magnetic moment, $G$'s denoting the familiar
Sachs form factors which are dimensionless and normalized to unity
at $Q^2=0$. For large $Q^2$, as a consequence of the $Q^2$ behavior of the r.h.s.
of (\ref{el}), we get
\be
\int^1_0 g_2(x,Q^2) dx=0.
\ee

In particular, from Eq.(9) it follows that
\be
I_2(0)={\mu_A^2+\mu_A e \over 4},
\ee
$e$ being
the nucleon charge in elementary units.
To reproduce the GDH value (see Eq.(5)) one should have
\be
I_{T}(0)={\mu_A e \over 4},
\ee
which was indeed proved by Schwinger \cite{Sch}.
The importance of the $g_2$ contribution can be seen
already, since the entire $\mu_A$-term for the GDH sum rule is
provided by $I_2$.

Note that $I_{T}$ does not differ from $I_1$ for large $Q^2$ due to the
BC sum rule, but it is {\it positive} in the proton case.
It is possible
to obtain a smooth interpolation for  $I^p_{T}(Q^2)$
between large $Q^2$ and $Q^2=0$ \ci{ST93}.

\be
I^p_{T}(Q^2)=\theta(Q^2_0-Q^2)({\mu_{A,p} \over 4}- {2 M^2 Q^2\over
{(Q^2_0)^2}} \Gamma^p_1)+\theta(Q^2-Q^2_0) {2 M^2\over {Q^2}}
\Gamma^p_1,
\ee
where $\Gamma^p_1=\int^1_0 g^p_1(x) dx$.
The continuity of the function and of its derivative is guaranteed
with the choice $Q^2_0=(16M^2/\mu_{A,p}) \Gamma^p_1 \sim 1GeV^2$,
where the integral is given by the world average proton data.
It is quite reasonable to distinguish
the perturbative and the non-perturbative regions. As a
result one obtains a crossing point at $Q^2 \sim 0.2 GeV^2$, below the
resonance region \ci{ST93}, while the positive value at $Q^2=0.5 GeV^2$
is in a good agreement with the E143 \ci{E143} data. 
A fair agreement with HERMES data has been also 
observed for larger $Q^2$ values \ci{HER}.     

This smooth interpolation seems to be very reasonable in the
framework of the QCD sum rules method as well. Then one should choose  some
"dominant" tensor structure to study the $Q^2$-dependence of its scalar
coefficient and $T_1$ appears to be a good candidate. This seems
also promising  from another point of view. It is not trivial to
obtain, within the QCD sum rules approach, the  GDH value at $Q^2=0$.
Since the r.h.s. of Eq.(12) is linear in $\mu_A$, it may be possible to
obtain it using the Ward identities, just like the normalization
condition for the pion form factor \ci{Rad}. This, in turn opens
the possibility to apply the powerful tool of quark-hadron duality.
The latter relies on the perturbative theory, and it is quite clear
that it is much more plausible to describe linear (one-loop), than quadratic
terms. One should recall here that while the sum rule for $I_{T}$ was checked
in QED long ago \ci{Mil}, the GDH sum rule required much more work
\ci{GDHQED}, since it gets non-trivial contributions only at two loops order,
although the vanishing value at one loop level is also non-trivial \ci{Alt,BS}.

Note also that the large contribution of $g_2$ by no means contradicts
the resonance approaches \ci{Ioffe} and may be
considered complimentary to them. In these cases, $\Delta(1232)$ plays
a central role: it provides a significant amount
of the GDH integral at $Q^2=0$ and gives a clear qualitative explanation
of rapid $Q^2$-dependence \ci{book}. The $\Delta$ photoproduction is
dominated by the magnetic dipole form factor, leading to 
a positive $I_2$ and a vanishing $I_T$, so implying a negative
$I_1$. The sign change is just related  to the fast decrease of the
$\Delta$ contribution \footnote{ For a more detailed discussion of
this important point see Ref.\ci{ST95}.}.

To generalize our approach to the neutron case, one needs a
similar smooth parametrization of $g_T$
for the neutron. Since the value at
$Q^2=0$ is equal to zero, it is not sufficient to limit oneself
to the simplest linear parameterization. One needs to
add a term, quadratic in $Q^2$. A simple parametrization
providing the continuity of the function and its derivative
was suggested in \ci{ST97}, which however, leads to a result
in contradiction with the data \cite{ZEM}.
This does not seem to be fortuitous, bearing in mind the
argument presented above. Indeed, the general reason
supporting the smoothness of interpolation for $g_T$ is
its linearity in $\mu_A$. As soon as this term appears to be equal to
zero for some special reason (which, in our case, is nothing but
than the neutron neutrality!), there is also no more reason to expect such
a smoothness.

To bypass this difficulty, we use the difference between proton and
neutron instead of the neutron itself. Although it is possible, in principle,
to construct a smooth interpolation for the functions $g_1$ themselves
\ci{book}, it does not fit the suggested general argument, since 
$I^{p-n}_{1}(0)$ is proportional to $\mu_{A,n}^2- \mu_{A,p}^2$, which
is quadratic and, moreover,
has an additional suppression due to the smallness
of isoscalar anomalous magnetic moment.

So we suggest the following parametrization for the isovector
contribution of $I_{T}(Q^2)$, namely $I^{p-n}_{T}(Q^2)$

\be
I^{p-n}_{T}(Q^2)=\theta(Q^2_1-Q^2)({\mu_{A,p} \over 4}- {2 M^2 Q^2\over
{(Q^2_1)^2}} \Gamma^{p-n}_1)+\theta(Q^2-Q^2_1) {2 M^2\over {Q^2}}
\Gamma^{p-n}_1,
\ee
where the transition value
$Q_1^2$ may be determined by the continuity
conditions in a similar way. We get the
value $Q_1^2 \sim 1.3 GeV^2$, which is of the same order
as for the proton case.

The elastic contribution to the BC sum rule should be included
for the neutron separately, so
we need
the neutron elastic form factors. While the electric one, might be
neglected, the magnetic form factor
is well described by the dipole formula \cite{FF}

\be
G_M (Q^2)={1 \over {(1+Q^2/0.71)^2}}.
\ee

The plot representing $\Gamma^n_1 (Q^2)$ is displayed
on Fig.1.

\begin{figure}
\begin{center}
\leavevmode {\epsfysize= 9cm \epsffile{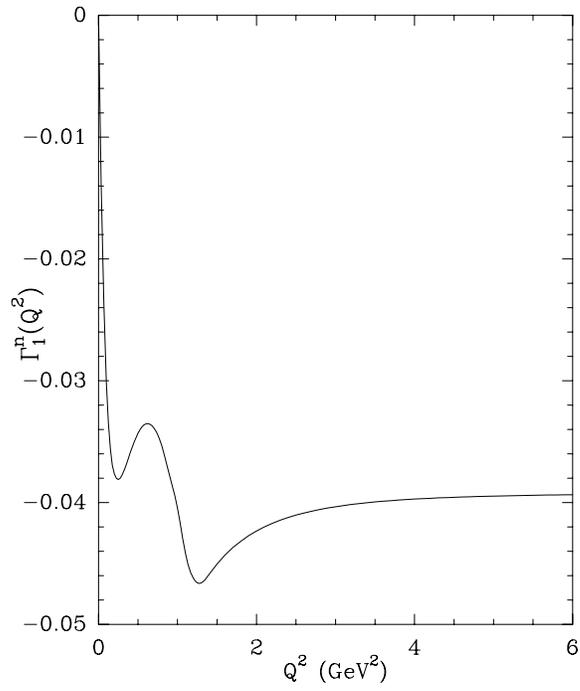}}
\end{center}
\caption[*]{\baselineskip 1pt
Our prediction for $\Gamma_1^{n}(Q^2)$.}
\label{Fig1}
\end{figure}

One can see, that $\Gamma^n_1 (Q^2)$ remains quite close to its asymptotic value
down to $Q^2$ values, as low as $3GeV^2$. Moreover, the preliminary data \cite{ZEM}
might bear some aspects of the structure generated in our approach,
by the interplay of the transitions values $Q_0^2$ and $Q_1^2$.

Now we have all the ingredients to show the behavior of $I_1^{p-n}(Q^2)$, a quantity 
directly related to the Bjorken sum rule $\Gamma_1^{p-n}(Q^2)$ (see Fig.2). 
It is clear that the smooth linear interpolation at low $Q^2$ to the value 
$(\mu_{A,n}^2- \mu_{A,p}^2)/4$, would 
result in a much lower value for $I_1^n(Q^2)$, inconsistent with the experimental 
data on $\Gamma^n_1 (Q^2)$ \cite{ZEM}.

\begin{figure}
\begin{center}
\leavevmode {\epsfysize= 9cm \epsffile{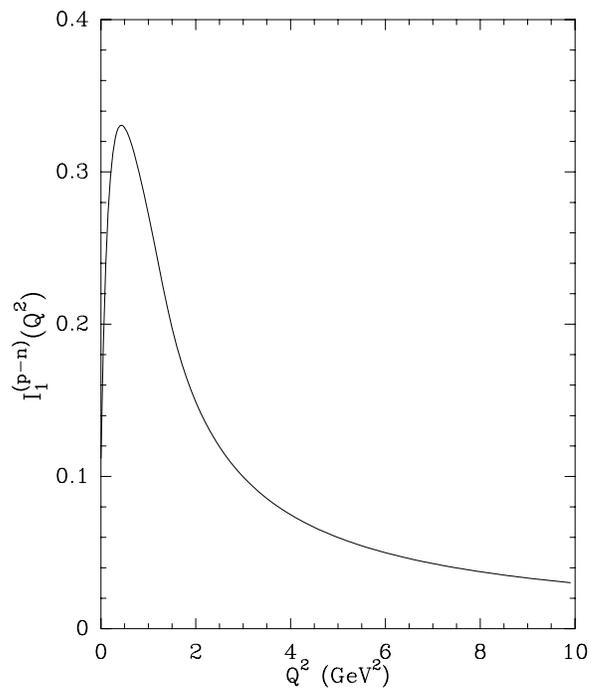}}
\end{center}
\caption[*]{\baselineskip 1pt
Our prediction for $I_1^{p-n}(Q^2)$, directly related to the Bjorken sum rule.}
\label{Fig2}
\end{figure}

Note that very accurate data from JLAB may require more
elaborate models for $g_{T}$. In particular, it is possible to take
into account the perturbative and power corrections to $I_1$, and also
the contribution of $I_2$ at the matching point $Q_0^2$, since in fact
we have assumed that at this point $I_T(Q_0^2)=I_1^{asymptotic}(Q_0^2)$.
However, these effects are acting in opposite directions
and should partially cancel each other. Indeed, in our model
$I_1(Q_0^2)=I_T(Q_0^2)-I_2(Q_0^2)=I_1^{asymptotic}(Q_0^2)-I_2(Q_0^2)$, and
$I_2$ produces a negative contribution, just like the corrections to $I_1$.

Another interesting problem is a possible violation
of the BC sum rule reported recently \ci{Rock}.
In this connection, note that from a theoretical point
of view, the only reason for a violation is the divergence of the integral.
So to put it in a dramatic manner,
"the BC integral is either zero or infinity". In this sense, the
reported finite value \ci{Rock} requires an interpretation.

One reason for the BC sum rule violation is the contribution of Regge cuts
\ci{book}, which are located at very low $x$ \ci{NNN}.
 Therefore, one may ask if it is possible to separate this contribution
\be
\label{sep}
g_2(x)= g_2^{BC}(x) + g_2^{cut}(x)
\ee
in such a way that
\be
\int^1_0 g_2^{BC}(x) dx=0.
\ee
Then one could define the
BC integral with a lower cutoff
\be
I_{\delta}=\int^1_{\delta} g_2(x) dx,
\ee
such that $I_{\delta}$ may be sufficiently close to zero
for small $\delta$, where $g_2^{cut}$ is still negligible.
We have considered such a separation \ci{ST95} and we have showed that
the crossing point for proton is {\it not} sensitive to the cut contribution.
Moreover, the details of $I_1$ may be sensitive to the
cut contribution and provide an additional way of its investigation.\\

{\bf Acknowledgments}

We are indebted to Claude Bourrely for some discussions. This research
was partly supported by INTAS (International Association for the Promotion of
Cooperation with Scientists from the Independent States of the Former Soviet
Union) under Contrat No. 00-00587.

\bb{99}
\bi{Bj} J. D. Bjorken, Phys. Rev. {\bf 148}, 1467 (1966);
Phys. Rev. {\bf D1}, 1376 (1970).
\bi{Pert} S.~A.~Larin, F.~V.~Tkachov and J.~A.~Vermaseren,
Phys.\ Rev.\ Lett.\  {\bf 66}, 862 (1991);
S.~A.~Larin and J.~A.~Vermaseren,
Phys. Lett.  {\bf B259}, 345 (1991).
\bi{Power} V.~M.~Braun and A.~V.~Kolesnichenko,
Nucl. Phys. B {\bf B283}, 723 (1987).
\bi{Ger} S. B. Gerasimov, Yad. Fiz. {\bf 2}, 598 (1965)
[Sov. J. Nucl Phys. {\bf 2}, 430(1966)].
\bi{DH} S. D. Drell and A. C. Hearn, Phys. Rev. Lett. {\bf 16}, 908 (1966).
\bi{book} B. L. Ioffe, V. A. Khoze and L. N. Lipatov, {\it Hard Processes},
(North-Holland, Amsterdam, 1984).
\bi{E143} E143Collaboration, K. Abe et al., Phys. Rev. Lett. {\bf78}, 815 (1997)
and Phys. Rev. {\bf D58}, 112003 (1998).
\bi{HER} HERMES Collaboration, A. Airapetian et al., Phys.Lett. {\bf B494}, 1 (2000).
\bi{ZEM} Z.-E. Meziani, Nucl. Phys. B (Proc. Suppl.) {\bf 105}, 105 (2002).    
\bi{ST93} J. Soffer and O. Teryaev, Phys. Rev. Lett. {\bf 70}, 3373 (1993).
\bi{ST95} J. Soffer and O. Teryaev, Phys. Rev. {\bf D51}, 25 (1995).
\bi{ST97} J. Soffer and O. Teryaev, Phys. Rev. {\bf D56}, 7458 (1997).
\bi{BC} H. Burkhardt and W. N. Cottingham, Ann. Phys. (N.Y.) {\bf 16}, 543 (1970).
\bi{Sch} J. Schwinger, Proc. Nat. Acad. Sci. U.S.A. {\bf 72}, 1559 (1975).
\bi{Rad}  V. A. Nesterenko and A. V. Radyushkin, Phys. Lett. {\bf B115}, 410 (1982).
\bi{Mil} Wu-Yang Tsai, L. L. DeRaad, Jr., and K. A. Milton, Phys. Rev. {\bf D11}, 3537 (1975).
\bi{GDHQED} D. A. Dicus and R. Vega, Phys. Lett. {\bf B501}, 44 (2001).
\bi{Alt} S. B. Gerasimov and J. Moulin, JINR E2-6722, 1972;
G. Altarelli, N. Cabibbo and L. Maiani, Phys. Lett. {\bf B40}, 415 (1972).
\bi{BS} S. J. Brodsky and I. Schmidt, Phys. Lett. {\bf B351}, 344 (1995).
\bi{Ioffe} V. D. Burkert and B. L. Ioffe, Phys. Lett. {\bf B296}, 223 (1992) and
J.\ Exp.\ Theor.\ Phys.\  {\bf 78}, 619 (1994) [Zh. Eksp. Teor. Fiz. {\bf 105}, 1153 (1994)];
V.~D.~Burkert, arXiv:nucl-ex/0109004.
\bi{FF}  A. Lung et al., Phys. Rev. Lett. {\bf 70}, 718 (1993).
\bi{Rock} E155 Collaboration, P. L. Anthony et al., arXiv:hep-ex/0204028.
\bi{NNN} I.~P.~Ivanov et al., Phys. Lett. {\bf B457}, 218 (1999).
\eb

\end{document}